\documentclass[prl,twocolumn,showpacs]{revtex4}
\usepackage{amsmath,bbm,epsfig}
\usepackage{amsfonts}
\usepackage{mathrsfs}
\newcommand{\mb}[1]{{\rm #1}}
\newcommand{\ii}{\mb{i}}
\begin{document} 

\title{Resonance-assisted decay of nondispersive wave packets}

\author{Sandro Wimberger,$^1$ Peter Schlagheck,$^2$ 
Christopher Eltschka,$^2$ and Andreas Buchleitner$^3$}  
\affiliation{$^1$CNR-INFM and Dipartimento di Fisica E. Fermi,
Unversit{\`a} degli Studi di Pisa, Largo Pontecorvo 3, 56127 Pisa, Italy \\
$^2$Institut f{\"u}r Theoretische Physik,
Universit{\"a}t Regensburg, 93040 Regensburg, Germany \\
$^3$Max-Planck-Institut f{\"u}r Physik komplexer Systeme, 
N{\"o}thnitzer Str. 38, 01187 Dresden, Germany}


\begin{abstract}
We present a quantitative semiclassical theory for the
decay of nondispersive electronic wave packets in driven, ionizing Rydberg 
systems. Statistically robust quantities are extracted combining resonance assisted tunneling with subsequent
transport across chaotic phase space and a final
ionization step.  
\end{abstract}

\pacs{03.65.Xp,32.80.-t,05.45.Mt,05.60.Gg}

\maketitle

The accurate and quantitative modeling of open
quantum systems at high spectral densities remains a serious challenge for  
quantum theory and computational physics. Even the ab initio treatment
of apparently simple, 
atomic one or two electron systems under nonperturbative forcing, in
the energy range of laboratory experiments, has
become available only during the last two decades
\cite{iu91}. Furthermore, it appears 
a safe bet that the driven three body Coulomb problem
\cite{scrinzi97} will be the 
most complicated ``many particle'' quantum problem still amenable to a
(numerically) exact and complete solution, even on the most powerful 
supercomputers. Thus,
alternative theoretical strategies which allow a precise treatment,
without saturating our computational resources nor loosing 
quantitative predictive power, are in need.

Specifically in the range of high spectral densities, semiclassical
approaches -- which try to deduce the spectral structure of a given
quantum system from the underlying Hamiltonian dynamics -- 
open such an avenue. However, while semiclassics of bounded
systems is well developed \cite{ozorio88}, 
its generalization for open systems \cite{garciamata04} 
has not been fully accomplished yet: Our semiclassical understanding of
tunneling and decay phenomena in systems with mixed regular-chaotic
classical phase space structure remains rather incomplete, and semiclassics
still has to prove its potential to come up with robust (a priori)
quantitative 
predictions for specific experiments. 

\begin{figure}
  \centerline{\epsfig{figure=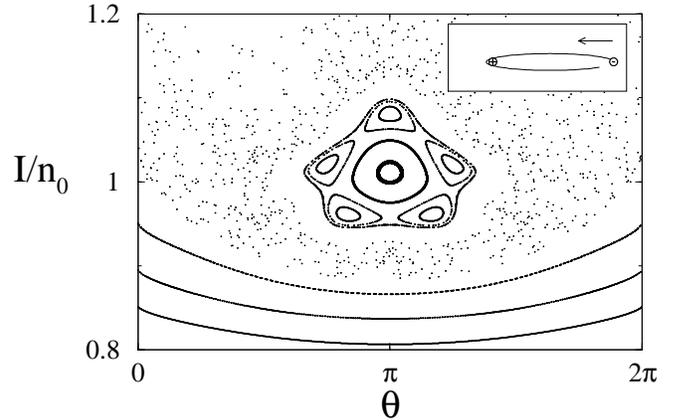,width=\linewidth}} 
  \caption{
    Classical phase space of driven hydrogen, generated by the
    Hamiltonian (\ref{eq:H}), for $\omega_0 = 1$ and 
    $F_0=0.041$,
    and spanned by the action-angle variables of the unperturbed atom, at 
    $\omega t = 0$.
    Nondispersive quantum wave packets are localized at the 
    center of the principal elliptic
    island (concentric structure in the middle of the plot) and move
    along the resonantly driven Kepler orbit sketched 
    in the inset (the arrow shows the direction of the driving force at the
    outer turning point), in phase with the driving field.
    For the chosen parameters, a prominent, higher order 
    nonlinear $5:1$ resonance strongly affects the tunneling 
    coupling of the principal 
    island's eigenstates  
    to the surrounding chaotic sea. This resonance is 
    expressed by the five-fold
    structure right within the outermost torus which confines the
    principal island.
  }
  \label{fig:1}
\end{figure}

In the present Letter, we improve on that situation, by elaborating a
fully quantitative semiclassical treatment of a 
paradigmatic example of open system dynamics at high spectral
densities, in 
the context of light-matter interaction and coherent control: 
Nondispersive
wave packets have been shown to be ubiquitous in periodically driven quantum
systems with underlying mixed regular-chaotic phase space 
\cite{arbo03,wimberger03,abu02,ibb94}. 
They can be launched along essentially arbitrary phase space trajectories, 
which can be manipulated in real time \cite{maeda05}. 
In contrast to ``traditional'' nondispersive wave packets, which rely on the
harmonicity of the spectra they are built on \cite{raman97},
these strongly localized eigenstates of the driven system are {\em robust}
against perturbations \cite{wimberger03,maeda04a,maeda05}, what is a
consequence of the Kolmogorov-Arnold-Moser theorem \cite{ozorio88}.
The characteristic property
which defines their unique 
potential for quantum control purposes (including quantum memory
applications \cite{ahn00}, i.e., within the framework of molecular
quantum computing \cite{tesch02}) is their essentially 
{\em eternal} life time on experimentally relevant time scales 
\cite{abu02,delande94}. 
The life time is determined by
the decay rate of the wave packet from the elliptic island to which it is
anchored to in classical phase space (see Fig.~\ref{fig:1}), i.e.,
it is essentially limited by a finite quantum mechanical tunneling
rate through classically impenetrable phase space barriers 
\cite{abu02,delande94,kuba98}. 

However, while always very long, the precise value of the wave
packet's life time is not a statistically robust
quantity -- it fluctuates over several orders of magnitude
under tiny changes of the boundary conditions, which may be stray
fields or other uncontrolled imperfections in the 
experiment. This sensitivity stems from the coupling of the wave
packet to the chaotic component of phase space, into which
it is completely embedded (see Fig.~\ref{fig:1}). We will
derive a quantitative semiclassical 
estimate for the mean value and the variance of the wave packet's
decay rate, without adjustable parameters. 

For the sake of clarity, we consider a
simplified (yet reliable \cite{abu02}) formulation of the problem we
have in mind: 
Atomic hydrogen initially prepared in 
a highly eccentric (extremal parabolic) Rydberg state 
aligned along the field polarization axis
of a linearly polarized microwave driving field of amplitude $F$ and frequency
$\omega$ can be described by the Hamiltonian 
\begin{equation}
  H(p_z,z) = \frac{p_z^2}{2} - \frac{1}{z} + F z \cos(\omega t)\, ,
  \label{eq:H}
\end{equation}
where the configuration space of the electron has been confined to the
$z$-axis (for the actual three dimensional problem, such confinement can be achieved
by applying a weak additional static electric field along the polarization
axis of the periodic drive \cite{abu02}). 
The classical phase space structure generated by this
Hamiltonian is illustrated in Fig.~\ref{fig:1}, and is completely fixed
by the choice 
of the scaled driving field frequency $\omega_0 = \omega n_0^3$ 
and amplitude $F_0=Fn_0^4$, respectively, 
as a consequence
of the scale invariance of Hamilton's equations of motion.
Specifically, we set $\omega_0=1$ and $F_0=0.041$, in order to realize the
essential ingredients of our subsequent treatment: (i) an elliptic island (at
the center of Fig.~\ref{fig:1}), which is (ii) completely embedded into the
chaotic sea, and (iii) encloses a prominent, higher order $r:s$ nonlinear
resonance right within its outermost confining torus.  

For
different values of $n_0$, and accordingly adjusted $\omega$ and $F$, we
find (by exact numerical diagonalization of the Floquet Hamiltonian
defined by (\ref{eq:H}) \cite{abu02}) wave packet eigenstates
localized on the elliptic island, together with 
their 
decay rates $\Gamma_{\rm wp}$. Under changes of $n_0$,
the latter
(which, in this specific physical realization, are nothing but
ionization rates mediated by multiphoton coupling to the atomic
continuum) fluctuate wildly, as shown in Fig.~\ref{fig:2} below. 

To determine the mean value and the variance of $\Gamma_{\rm wp}$,
let us start from the random matrix ansatz for an effective Hamiltonian 
employed in  \cite{kuba98}:      
\begin{equation}
  H_{\mb{eff}} = \left( 
    \begin{array}{ccccc}
      E_{\rm wp} & V_{\rm rc} & 0 & \cdots & 0 \\
      V_{\rm rc} & H_{11} & \cdots & \cdots & H_{1N} \\
       0 & \vdots & & & \vdots \\
       \vdots &  H_{N1} & \cdots & \cdots & H_{NN} - \frac{\ii}{2} \Gamma_c
    \end{array}
  \right) \, .
  \label{eq:model}
\end{equation}
The matrix structure comprises the wave packet state localized 
at the center of the regular island (with energy $E_{\rm wp}$), 
and $N$ additional states located in the chaotic phase space region,
one of which being subject to decay with the rate $\Gamma_c$.
The chaotic states are assumed to be strongly coupled to each other, such 
that the corresponding sub-block $(H_{ij})$ can be described by a random
matrix from the Gaussian orthogonal ensemble \cite{TU1994}.
In accordance with the resonance-assisted coupling mechanism to be
explained below, we assume that the wave packet is dominantly 
coupled to only one of those chaotic states, via the (comparatively small) 
matrix element $V_{\rm rc}$. 
As a qualitative, conceptional improvement over \cite{kuba98},
where $V_{\rm rc}$ and $\Gamma_c$ were free fitting parameters, we
here derive quantitative estimates for these two coupling constants.

Within our semiclassical framework, transport to the continuum is 
only possible by traversing a large part of chaotic phase
space. Quantum mechanically, this is equivalent to multiphoton
transition amplitudes which couple the initial state -- the wave
packet -- to Rydberg states with ionization potentials smaller than 
$\omega$, i.e., with principal quantum numbers $n$ above 
$n_c = n_0^{3/2} / \sqrt{2 \omega_0}$.
The final, ionizing step occurs from these highly excited states by a
one-photon process. 
In the effective Hamiltonian of (\ref{eq:model}), this final 
ionization process is accounted for by attributing a finite decay
rate $\Gamma_c$ to one of the chaotic states, which is
well estimated by the Golden Rule expression
\begin{equation}
  \Gamma_c = 0.265 F^2 \omega^{-10/3}n_c^{-3} \simeq 1.26 \times
  10^{-3} n_0^{-5/2} {\rm a.u.} 
  \label{eq:gamma}
\end{equation}
for the single-photon ionization rate of a Rydberg state with quantum number
$n_c$ \cite{casati88}.

A random-matrix average over the
eigenvalues and eigenvectors of the chaos block $(H_{ij})$ gives rise to the
Cauchy-type probability density 
\begin{equation}
  P(\Gamma_{\rm wp}) = \frac{1}{\pi} \frac{\sqrt{\Gamma_0 /
  \Gamma_{\rm wp}}}{\Gamma_{\rm wp} + \Gamma_0} 
  \label{eq:prob}
\end{equation}
of the wave packet's decay rate \cite{kuba98}. 
The characteristic scale $\Gamma_0$ is derived as $\Gamma_0 = (\pi
V_{\rm rc}/\omega)^2  \Gamma_c$, 
exploiting that the chaotic eigenvalues are uniformly distributed
within an energy interval of width $\omega$, due to the inherent
periodicity of the Floquet spectrum. 
Because of the large fluctuations implied by the distribution
(\ref{eq:prob}), the relevant statistical quantity is given by 
the averaged logarithm \cite{kramer93}. 
An explicit calculation of this {\em geometric} mean gives
\begin{equation}
  \overline{\Gamma}_{\rm wp} \equiv 
\exp[\langle\ln(\Gamma_{\rm wp})\rangle] = \Gamma_0 = 
  \left( \frac{\pi V_{\rm rc}}{\omega}\right)^2 \Gamma_c\, ,
\label{eq:avergamma}
\end{equation}
which depends only on the single-photon ionization rate $\Gamma_c$ from
(\ref{eq:gamma}), on the driving 
frequency $\omega$, and on the coupling matrix element $V_{\rm rc}$ between the
regular island and its chaotic surrounding. $V_{\rm rc}$ can be derived using
purely semiclassical arguments, as follows.

Let us recall that nonlinear resonances between the
external drive and the local modes of a regular phase space region
induce {\em higher order} perturbative couplings between the locally
quantized eigenstates of 
that region \cite{ozorio84}.
This leads to the phenomenon of \emph{resonance-assisted tunneling}
  which, originally proposed for near-integrable dynamics, can be generalized
  to mixed regular-chaotic systems as well \cite{eltschka05}.
  There it provides the dominant semiclassical mechanism for the tunneling
  process that connects the ``ground state'' of a regular island (localized at
  the island's center) to states within the chaotic sea (which, in
  turn, are coupled to the atomic continuum with a rate $\Gamma_c$, given
  in Eq. (\ref{eq:gamma})).

Quantitatively, the classical dynamics near a $r$:$s$ resonance ($s$
oscillations match $r$ driving periods) is described by the effective pendulum
Hamiltonian \cite{ozorio84}
\begin{equation}
  H_{\rm res} = \frac{(\widetilde{I} -I_{r:s})^2}{2m_{r:s}} + 
  2 V_{r:s} \cos( r \widetilde{\theta} )\, , \label{eq:pend}
\end{equation}
expressed in terms of the local action-angle variables
$\widetilde{I},\widetilde{\theta}$ of the principal elliptic 
island, where $s$ is
absorbed in the definition of $\widetilde{\theta}$.
The central action $I_{r:s}$ of
the $r$:$s$ 
resonance, and the effective mass and coupling parameters $m_{r:s}$, 
$V_{r:s}$, can be extracted with little numerical effort from the 
classical phase space, by computing the area covered by the separatrices of
the $r$:$s$ resonance, and by evaluating the stability matrix of the associated
periodic points \cite{eltschka05}.
Quantizing (\ref{eq:pend}), we see that the perturbation term induced by the
$r$:$s$ resonance couples the $k$th excited state of the principal island,
given by the plane wave $\langle\widetilde{\theta}|k\rangle \sim
\exp(i k \widetilde{\theta})$ in 
angle space, to the states 
$|k \pm r\rangle$, with a strength $V_{r:s}$. Since the wave
packet state corresponds to the principal island's ground state with
$k=0$, its coupling to the chaotic sea is thus given by the 
effective matrix element 
\begin{equation}
  V_{\rm rc} = V_{r:s} \prod_{l=1}^{l_c-1} \frac{V_{r:s}}
  {\widetilde{E}_0 - \widetilde{E}_{l \cdot r}}\, , \label{eq:vrc}
\end{equation}
with $\widetilde{E}_{k} = (\widetilde{I}_k -I_{r:s})^2 
/ (2 m_{r:s})$ and $\widetilde{I}_k = \hbar (k + 1/2)$. Here,
$|l_c \cdot r\rangle$
denotes the first state within 
this perturbative sequence of higher order couplings 
that is located outside the principal 
island, and the prefactor $V_{r:s}$ accounts for the final step
from $|(l_c-1)\cdot r\rangle$ to $|l_c \cdot r\rangle$.

In our exemplary case of Fig.~\ref{fig:1}, the driving 
induces a prominent 5:1 resonance within the principal regular island on which
the wave packet is localized. 
From the area covered by the principal island -- which is determined
numerically 
-- we deduce  
$n_0 \simeq 135$ as the critical quantum number at which the island's fifth
excited state 
(i.e., the first state to which the ground state is coupled by the 5:1
resonance) is located exactly on the outermost invariant torus.
For $n_0 < 135$, we can therefore identify this fifth excited state with the
first basis state of the chaos block in the Hamiltonian (\ref{eq:model}), and
set $V_{\rm rc} = V_{r:s}$, which, together with
Eqs.~(\ref{eq:gamma},\ref{eq:avergamma}) and the numerically computed
value of $V_{r:s}$, leads to our central result
\begin{equation}
  \overline{\Gamma}_{\rm wp} \simeq 9.6 \times 10^{-13} n_0^{-1/2}
  {\rm a.u.}\, ,\, n_0<135,
  \label{eq:gamma2}
\end{equation}
for the mean decay rate of the wave packet.
At larger principal quantum numbers ($n_0 > 135$), the coupling from the
regular island to the chaotic sea 
involves one more perturbative step in
Eq.~(\ref{eq:vrc}), which substantially reduces the matrix element
$V_{\rm rc}$.
This leads to
\begin{equation}
  \overline{\Gamma}_{\rm wp} \simeq 1.7 \times 10^{-20} \frac{n_0^{7/2}}{(n_0 - 131)^2}
  {\rm a.u.}\, ,\, n_0 > 135\, .
  \label{eq:gamma3}
\end{equation}
As a consequence, the semiclassically calculated decay rates exhibit a sudden
drop at $n_0 \simeq 135$. (\ref{eq:gamma3}) describes the monotonic decrease
(mind the expression's denominator!)
of the rates with $n_0$ up to $n_0 \simeq 250$, where an additional drop 
is expected to occur by virtue of yet another perturbative step in 
Eq.~(\ref{eq:vrc}).
\begin{figure}
  \centerline{\epsfig{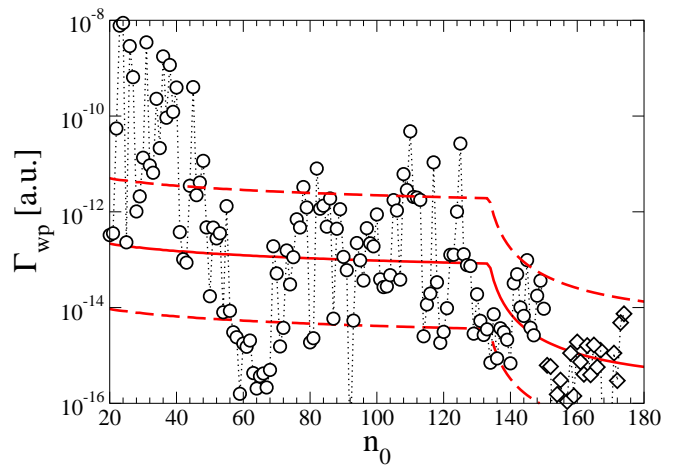}} 
  \caption{(color online). 
    Comparison of the exact quantum decay rates of the wave packet states
    (symbols) with the semiclassical prediction for their mean value 
    $\overline{\Gamma}_{\rm wp}$ 
    [solid line, given by Eqs.~(\ref{eq:gamma2}) and
    (\ref{eq:gamma3})] and the standard deviation [dashed lines, 
    see Eq.~(\ref{eq:sdev})].
    The rates are plotted as a function of the principal quantum number  
    $n_0$ of the Rydberg orbit along which the wave packet is launched.
    The quantum results were obtained for the same field parameters as
    Fig.~\ref{fig:1}. Open circles and diamonds distinguish different
    basis sizes form the numerical diagonalization. For $n_0\leq 150$,
    the basis was chosen such as to resolve all bound states
    with an ionization potential larger and equal to the one-photon
    energy $\omega$. For $n_0>150$, only bound states below $2n_0$
    were numerically resolved.
  }
  \label{fig:2}
\end{figure}

Fig.~\ref{fig:2} highlights that our semiclassical result 
$\overline{\Gamma}_{\rm wp}$ fits the exact quantum decay rates 
$\Gamma_{\rm wp}$ of the wave packet very well.
On average, after a first exponential decrease for 
$n_0 \leq 60$, the decay rate 
settles to a plateau at
$\overline{\Gamma}_{\rm wp} \simeq 10^{-13}$ a.u., and eventually 
drops again around $n_0 \geq 135$. Beyond that, our model also
succeeds to very well characterize the quantum rates' fluctuations
by their logarithmic standard deviation derived from (\ref{eq:prob}):
\begin{equation}
  \sqrt{{\rm var} [\ln(\Gamma_{\rm wp})]} = 
  \sqrt{\left\langle\left[\ln(\Gamma_{\rm wp}) - 
        \ln(\overline{\Gamma}_{\rm wp})\right]^2\right\rangle} = \pi \, .
  \label{eq:sdev}
\end{equation}
This {\em universal} result is indicated by the dashed lines in
Fig.~\ref{fig:2}, which are given by 
$\overline{\Gamma}_{\rm wp} e^{\pm \pi}$ and which encompass
the overwhelming part of the data.
Despite the sensitive dependence of individual decay
rates on boundary conditions and classical phase space structures,
our combined semiclassical and random matrix approach provides a
robust statistical characterization, over a wide range of excitation energies.

The disagreement between our semiclassical result and the quantum data below
$n_0 \simeq 60$ is not surprising, since, at such low
excitations,
the principal island cannot accommodate a sufficient number of quantum
eigenstates to resolve the $5:1$ resonance (approx. $r/2$ states 
are needed for satisfactory resolution \cite{eltschka05}). Thus, a necessary
condition (item (iii) above) for the resonance assisted tunneling 
mechanism to manifest itself is not satisfied at too small $n_0 \lesssim 60$.

In conclusion, the plateau-like structure born out in Fig.~\ref{fig:2}, in the range
$n_0\simeq 60\ldots 135$, corroborates the relevance of the resonance
assisted tunneling mechanism for the decay properties of nondispersive wave
packets, in an experimentally routinely accessible energy range
\cite{arbo03,maeda05}. Note that the average trend $\Gamma_{\rm wp}\sim
n_0^{-1/2}$, 
predicted by (\ref{eq:gamma2}) and confirmed in Fig.~\ref{fig:2}, is
markedly different from an on average exponential decay with $n_0$, as
assumed in \cite{kuba98,hornberger98}. 
  Indeed, a similar step structure arises in earlier published data on
  nondispersive hydrogen wave packets \cite{hornberger98} (at a slightly
  different field amplitude), which, on the basis of our presently
  improved understanding, is also attributed to resonance-assisted tunneling.
Thus, while the precise
extension of the plateau depends on the specific choice of parameter values,
and certainly also on the effective dimensionality of the electron's
configuration space, the structure as such is a robust fingerprint of
resonance assisted tunneling, as well as the universal value $\pi$ of the
decay rates' standard deviation. An experimental verification of this
prediction is challenging, though appears in reach for the most advanced
experimental setups to date \cite{maeda05,arbo03}. 

We thank Italo Guarneri for inspiring discussions, and acknowledge support
by the Alexander von Humboldt Foundation and the DFG.

\end{document}